# Journal of Geophysical Research: Space Physics

**RESEARCH ARTICLE**
10.1002/2015JA022058
# Dynamic effects of restoring footpoint symmetry on closed magnetic field lines

**Key Points:**
- Observed asymmetric convection and FAC on closed field lines consistent with effect of asymmetric stress release during IMF By conditions
- The event experiences a large (3 h) MLT displacement of the nightside aurora between the two hemispheres
- The restoring symmetry process can be an important mechanism in creating asymmetric FACs and convection on closed field lines

**Correspondence to:**
J. P. Reistad,
jone.reistad@uib.no

**Citation:**
Reistad, J. P., N. Østgaard, P. Tenfjord, K. M. Laundal, K. Snekvik, S. Haaland, S. E. Milan, K. Oksavik, H. U. Frey, and A. Grocott (2016), Dynamic effects of restoring footpoint symmetry on closed magnetic field lines, *J. Geophys. Res. Space Physics*, *121*, doi:10.1002/2015JA022058.

Received 20 OCT 2015
Accepted 21 MAR 2016
Accepted article online 28 MAR 2016
J. P. Reistad[1], N. Østgaard[1], P. Tenfjord[1], K. M. Laundal[1], K. Snekvik[1], S. Haaland[1,2], S. E. Milan[1,3], K. Oksavik[1,4], H. U. Frey[5], and A. Grocott[6]

[1]Birkeland Centre for Space Science, Department of Physics and Technology, University of Bergen, Bergen, Norway, [2]Max Planck Institute for Solar System Research, Göttingen, Germany, [3]Department of Physics and Astronomy, University of Leicester, Leicester, UK, [4]University Centre in Svalbard, Longyearbyen, Norway, [5]Space Sciences Laboratory, University of California, Berkeley, California, USA, [6]Physics Department, Lancaster University, Lancaster, UK
**Abstract** Here we present an event where simultaneous global imaging of the aurora from both hemispheres reveals a large longitudinal shift of the nightside aurora of about 3 h, being the largest relative shift reported on from conjugate auroral imaging. This is interpreted as evidence of closed field lines having very asymmetric footpoints associated with the persistent positive $y$ component of the interplanetary magnetic field before and during the event. At the same time, the Super Dual Auroral Radar Network observes the ionospheric nightside convection throat region in both hemispheres. The radar data indicate faster convection toward the dayside in the dusk cell in the Southern Hemisphere compared to its conjugate region. We interpret this as a signature of a process acting to restore symmetry of the displaced closed magnetic field lines resulting in flux tubes moving faster along the banana cell than the conjugate orange cell. The event is analyzed with emphasis on Birkeland currents (BC) associated with this restoring process, as recently described by Tenfjord et al. (2015). Using data from the Active Magnetosphere and Planetary Electrodynamics Response Experiment (AMPERE) during the same conditions as the presented event, the large-scale BC pattern associated with the event is presented. It shows the expected influence of the process of restoring symmetry on BCs. We therefore suggest that these observations should be recognized as being a result of the dynamic effects of restoring footpoint symmetry on closed field lines in the nightside.
## 1. Introduction

When the interplanetary magnetic field (IMF) interacts with the Earth's magnetic field, a dawn-dusk component in the IMF will affect the two hemispheres differently in a number of ways. The ionospheric convection pattern is commonly attributed to represent the footprints of this interaction and has been studied in great detail in both hemispheres [*Heppner and Maynard*, 1987; *Pettigrew et al.*, 2010; *Cousins and Shepherd*, 2010; *Förster and Haaland*, 2015]. It is known that the presence of IMF $B_y$ alters the usual two-cell convection pattern into a crescent "banana" and a round "orange" cell with locations approximately mirrored across the noon-midnight meridian in the two hemispheres. However, an outstanding challenge when comparing the two hemispheres is to resolve the true conjugate regions. Simultaneous global auroral imaging from both hemispheres has proven to be able to identify such connected regions as the aurora serves to "light up" the footprints of magnetospheric processes. A highly distinguishable feature in the nightside aurora, the substorm onset, has been studied in this way. Its hemispheric relative displacement in longitude has been shown to be largely controlled by IMF $B_y$ [*Liou and Newell*, 2010; *Østgaard et al.*, 2005, 2011b].

It has been shown by theory [*Cowley*, 1981; *Khurana et al.*, 1996], observations [*Cowley and Hughes*, 1983; *Lui*, 1984; *Wing et al.*, 1995], and modeling [*Kullen and Janhunen*, 2004; *Guo et al.*, 2014; *Tenfjord et al.*, 2015] that for nonzero IMF $B_y$ a $B_y$ component in the same direction as IMF $B_y$ is induced on closed field lines. Consistent with this, we observe asymmetric footpoints. Although there are other processes leading to $B_y$ in the closed magnetosphere, such as dipole tilt-related effects, $B_y$ induced from IMF $B_y$ is considered as the primary contributor in the plasma sheet [*Petrukovich*, 2011].

IMF $B_y$-related changes in the $B_y$ component in the closed magnetosphere is well understood and explained by considering the forces acting on the magnetosphere due to IMF interacting with the terrestrial field [*Cowley*, 1981; *Guo et al.*, 2014; *Tenfjord et al.*, 2015; *Østgaard et al.*, 2015]. The study by *Tenfjord et al.* [2015] emphasized that the commonly used terminology "IMF $B_y$ penetration" is misleading in this regard, as the changes in the $B_y$ field in the magnetosphere is not a simple fraction of the IMF $B_y$ nor can it be regarded as a vacuum superposition of external (IMF) field contributions. Using the Lyon-Fedder-Mobarry global magnetohydrodynamic (MHD) model, they showed that $B_y$ is induced in the closed magnetosphere as a result of the asymmetric loading of magnetic flux in the lobes during IMF $B_y$ conditions. This gives dawn-dusk asymmetric pressure in the two lobes. The plasma convection in the two lobes will be opposite and a $B_y$ component on closed field lines will be induced. This asymmetric forcing between the hemispheres also causes the footpoints to move relatively between the hemispheres, introducing a longitudinal displacement of conjugate points from a quiet time position. Throughout the manuscript we will use the same terminology as *Tenfjord et al.* [2015] for the appearance of $B_y$ in the magnetosphere due to IMF $B_y$, namely, *induced* $B_y$.

©2016. The Authors.
This is an open access article under the terms of the Creative Commons Attribution License, which permits use, distribution and reproduction in any medium, provided the original work is properly cited.
REISTAD ET AL.   EFFECTS OF RESTORING FOOTPOINT SYMMETRY   1

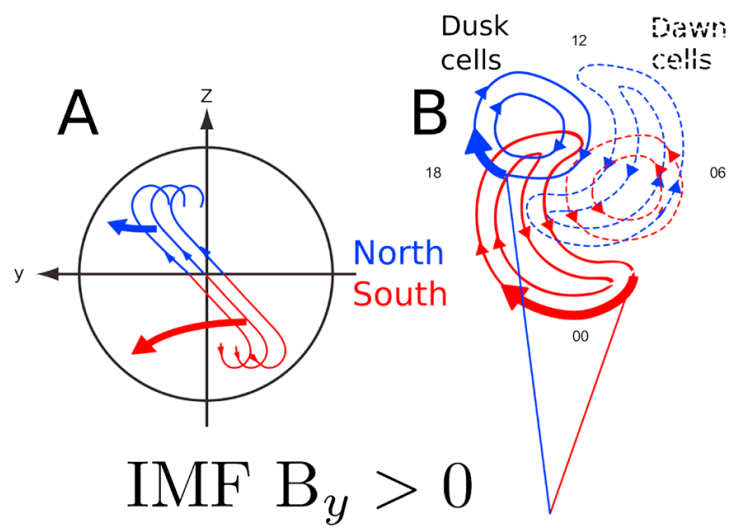

**Figure 1.** (a) View from deep in the magnetotail toward the Earth of magnetic field lines with displaced footpoints due to IMF $B_y$ positive. Arrows indicate the plasma flow toward the dayside in the dusk convection cells. (b) A view of how a magnetic field line in the dusk cells (seen in Figure 1a) is connected to the two hemispheres. The perspective is from above the northern magnetic pole looking through the Earth. The small numbers refer to MLT coordinates. Blue represent the Northern Hemisphere, and red represents the Southern Hemisphere. The dawn convection cells are also shown (dashed) but not with any connecting field line (after Figure 3 of *Grocott et al.* [2005]).

When illustrating the Earth's magnetic field lines with an induced $B_y$, as seen in Figure 1a, they will have asymmetric footpoints in the ionosphere [*Liou and Newell*, 2010; *Østgaard et al.*, 2011b; *Motoba et al.*, 2010]. The topic of this paper is the dynamic effects when these field lines gradually relax to a more symmetric situation as they move closer to Earth and toward dawn or dusk.

The dynamics of magnetic field lines on the nightside having an induced $B_y$ have been studied earlier with emphasis on the observed plasma convection in the magnetosphere and ionosphere. Although first suggested by *Nishida et al.* [1998], *Grocott et al.* [2004] were the first to observe the fast east/west convection in the nightside ionosphere on field lines supposedly having asymmetric footpoints. What *Grocott et al.* [2004] referred to as fast flows was ∼1000 m/s plasma convection in the midnight sector typically extending 2–3 h of magnetic local time (MLT) and its east/west direction determined by the sign of IMF $B_y$ and the hemisphere. It was later shown that these plasma flow signatures, indeed, were simultaneously present in the magnetosphere and ionosphere and oppositely directed in the two hemispheres [*Grocott et al.*, 2007]. This has later been confirmed by *Pitkänen et al.* [2015], also looking at simultaneous magnetospheric and ionospheric convection in both hemispheres, indicating that restoring footpoint symmetry is a fundamental process throughout the Dungey cycle, much like any physical system trying to restore a minimum energy configuration. It should be noted that the plasma flows seen by Grocott et al. were mostly occurring during northward IMF [*Grocott et al.*, 2008]. However, this mechanism should still work for southward IMF as the tail will still be asymmetrically forced by IMF $B_y$. One effect of restoring footpoint symmetry is schematically illustrated in Figure 1 (after Figure 3 of *Grocott et al.* [2005]) during positive IMF $B_y$ conditions. Figure 1a shows three field lines on the nightside with asymmetric footpoints, seen from the magnetotail, where blue color indicates the Northern Hemispheric end, and red is used for the Southern Hemispheric end. The field lines in Figure 1a are assumed to convect around the Earth in the dusk cell (since our event is focusing on the dusk cell). Hence, their magnetospheric and ionospheric velocities (large arrows in Figures 1a and 1b) point in that direction. Since the footpoints of the field line are displaced in longitude on the nightside, the Southern Hemispheric end needs to move faster in the westward direction in the dusk cell to catch up with the Northern Hemisphere as the field line convects toward the dayside. This is indicated in the ionospheric view in Figure 1b. Here one of the field lines from Figure 1a is shown to have asymmetric footpoints and also different westward convection speeds reflected by the different lengths of the thick arrows. In the dawn cell the situation is opposite (not shown here). These asymmetric convection speeds can be explained by considering the forces acting on the field line, shown toward the end of this section when we introduce Figure 2.

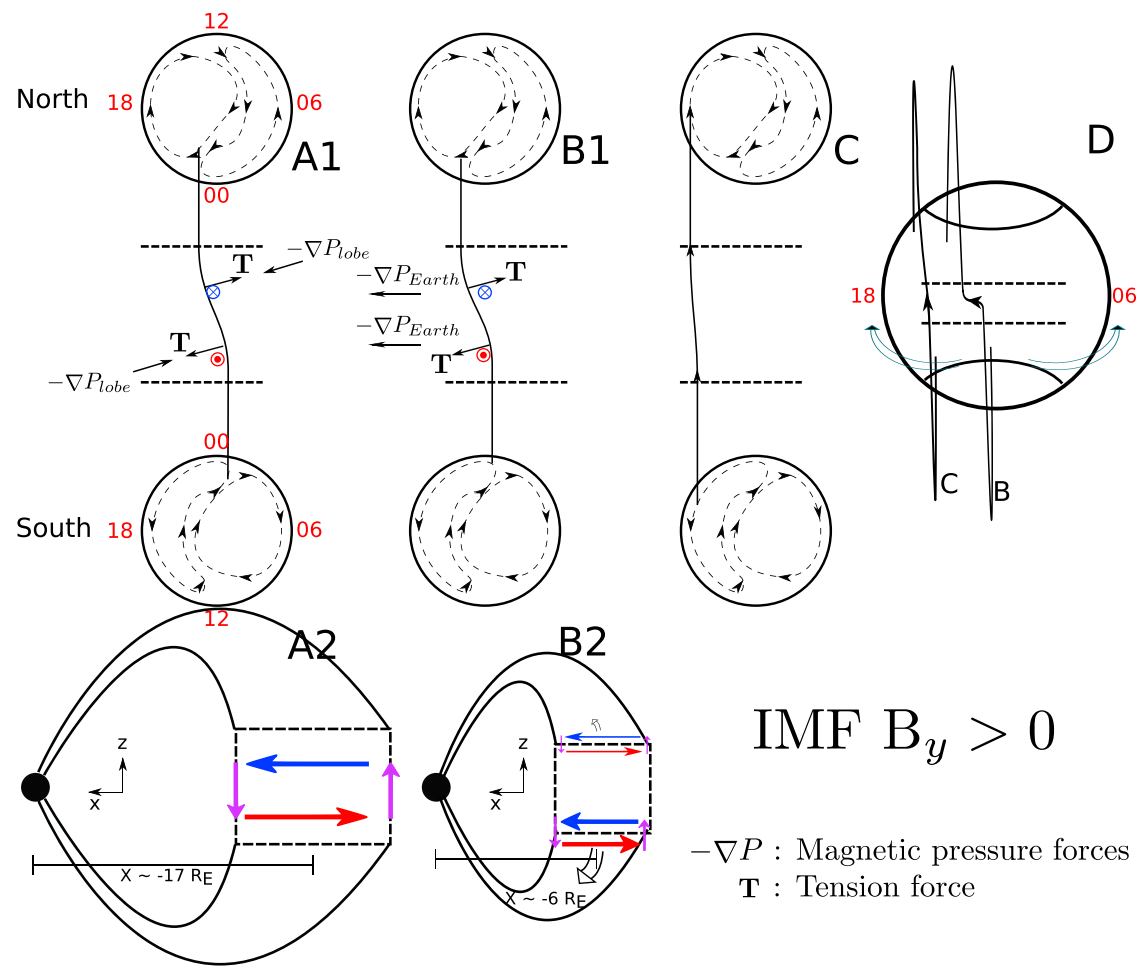

**Figure 2.** (a–c) Evolution of a closed field line with asymmetric footpoints in the dusk convection cells during IMF $B_y$ positive conditions. Figures 2a1, 2b1, and 2c show how a field line with asymmetric footpoints connects to the dusk convection cells in the two hemispheres at three different times. (a2 and b2) The associated current systems with colored arrows, seen from the dusk side. Figure 2a1: In the midtail region the asymmetric pressure forces due to IMF $B_y$ ($-\nabla P_{lobe}$) and the magnetic tension forces ($\vec{T}$) on the field line balance. Currents close locally as indicated in Figure 2a2. Figure 2b1: At a later stage the field line moves earthward and is affected by the gradient of the total pressure surrounding the Earth (plasma and magnetic field, $-\nabla P_{Earth}$). Now the forces do not balance. In the Southern Hemisphere the indicated forces point in the same direction. Hence, most of the stress is transmitted into this hemisphere, and the southern footpoint will catch up with the northern counterpart to restore symmetry, as seen in Figure 2c. (d) The situation in Figures 2b and 2c viewed from the tail.

The magnetospheric geometry for nonzero IMF $B_y$ has been widely investigated. *Stenbaek-Nielsen and Otto* [1997] suggested that the modified magnetic field geometry due to induced $B_y$ in the nightside (Figure 1) should give rise to interhemispheric currents due to the finite extent of the $B_y$ component in the $x$ direction (toward the Sun) in the magnetotail, explaining their observations of nonconjugate aurora. However, no explanation on how the currents could propagate all the way to the ionosphere was given. More recent observations of nonconjugate aurora [*Østgaard et al.*, 2011a; *Reistad et al.*, 2013] have used similar arguments to suggest that the field configuration due to the induced $B_y$ could be responsible for the observed asymmetries. However, the relation of such observations to the process of restoring symmetry was only recently proposed [*Tenfjord et al.*, 2015].

The view presented by *Grocott et al.* [2005] (our Figure 1) gives a conceptual framework for understanding the plasma convection observations, both from space and from the ground. However, its relation to field-aligned currents, also known as Birkeland currents (BCs), has not been treated in a unified way until recently [*Tenfjord et al.*, 2015]. The present study largely builds on this advancement in understanding. We present an event with a large longitudinal displacement of footpoints and apply the understanding of this configuration to explain the observed convection, aurora, and associated BCs in the framework of restoring footpoint symmetry.

Before we present our event we give a description of what we will refer to as the process of restoring symmetry, which takes place when a field line with asymmetric footpoints relaxes gradually to a more symmetric configuration as it moves toward dusk (or dawn). Figure 2 (after Figure 6 of *Tenfjord et al.* [2015]) shows a simplified model of the evolution (and related currents) of a field line on the nightside having asymmetric footpoints when considering the magnetospheric forces acting on it for the IMF $B_y$ positive case. As pointed out by *Tenfjord et al.* [2015], the induced $B_y$ in the tail is due to the buildup of an asymmetric magnetic pressure distribution in the two lobes as the tension forces on the newly reconnected field lines on the dayside act in opposite directions in the two hemispheres. For southward (but $B_y$ dominated) IMF, flux is added asymmetrically, and for northward IMF, flux is rearranged asymmetrically. In both cases one will get an asymmetric magnetic pressure distribution in the two lobes which cause asymmetric plasma convection within the magnetosphere and affect also the closed field lines differently in the two hemispheres. The shear plasma flows between the northern and the southern halves of the magnetotail will eventually also displace the footpoints of the field lines [*Liou and Newell*, 2010; *Guo et al.*, 2014; *Tenfjord et al.*, 2015].

The asymmetric pressure forces ($-\nabla P_{lobe}$ in Figure 2a1) act on the field line in opposite directions in the two hemispheres creating asymmetric footpoints. Due to this, there will be an oppositely directed tension force, $\vec{T}$ acting to balance the $-\nabla P_{lobe}$ force. The illustrated field line in Figure 2a1 is a closed field line in the midtail convecting toward the Earth and returning in the dusk cell (same situation as in Figure 1). How the field line connects to the banana and the orange cell in opposite hemispheres is also indicated. Associated with this geometry, Ampère's law requires a pair of currents flowing in the *x* direction as indicated by red and blue arrows in and out of the plane in the middle of Figure 2a1. In Figure 2a2 the field line is seen from the side. The box in the equatorial region indicates an idealized extent of the $B_y$ component in the *x* direction, meaning that $B_y$ is zero on either side. At these edges, $\nabla \times \vec{B}$ implies a current in the *z* direction as indicated by the purple arrows. In this case, when the forces balance, the currents close locally, and no BCs into the ionosphere are present.

As the field line convects closer to the Earth it will experience the total pressure from Earth, $-\nabla P_{Earth}$, and be less affected by $-\nabla P_{lobe}$, see Figure 2b1. At this location $-\nabla P_{Earth}$ has a component toward dusk along the entire field line since we have chosen to consider a field line returning in the dusk cell. In this situation the forces do not balance but are highly unbalanced as the forces are added in one hemisphere and subtracted in the other. The stress stored in the field will now be transported primarily toward the Southern Hemisphere where the forces are added. We emphasize that these considerations are based on an MHD description of how the system responds to external forcing.

Seen from the side, Figure 2b2 illustrates that this situation represents Alfvèn wavefronts carrying a pair of BCs that are launched primarily into the Southern Hemisphere. These waves act to restore footpoint symmetry along the equatorward part of the banana cell as the stress is being released. Due to this, the BCs transmitted (by the waves) to the Southern Hemisphere are expected to be stronger than in the other hemisphere. This illustrates a situation where asymmetric (between hemispheres) BCs can be generated. The net effect is the same as suggested by *Stenbaek-Nielsen and Otto* [1997]. However, this modified picture puts further constraints on the location of where the asymmetric BCs should be located (dawn or dusk cell). Figure 2c illustrates a later stage when the field line has relaxed to a more symmetric configuration. Figure 2d illustrates the transition from step B to C as viewed from the tail.

To summarize, when we refer to the process of restoring symmetry, we mean the following. (1) The release of magnetic stress from the magnetosphere to the hemisphere connected to the banana cell that will gradually restore the asymmetric field lines as they convect duskward or dawnward. (2) This is consistent with faster azimuthal ionospheric plasma flow along the banana cell than along the conjugate orange cell. (3) The asymmetric release of magnetic stress also leads to asymmetric Birkeland currents into the conjugate hemispheres. The term "untwisting process/phenomenon/hypothesis" that has been used in the literature [*Pitkänen et al.*, 2015] includes only point 2 of this process. To avoid misunderstanding, we will therefore use the term 'restoring symmetry process' throughout the paper.

In the following we present a case where simultaneous global auroral imaging reveals a large longitudinal displacement of the nightside aurora indicating the asymmetric footpoint configuration discussed above. Simultaneous conjugate ionospheric plasma flow measurements in the nightside convection throat region are also presented, allowing us to identify the extent of the convection cells in both hemispheres and compare with the auroral display. As no simultaneous BC measurements were available during the event, we investigate average BC maps obtained during similar conditions as the event to explore the impact of the process of restoring symmetry on BCs in the two hemispheres and compare with the model predictions from Figure 2. The emphasis will be on investigating how the observations fit our understanding of the large-scale electrodynamics in terms of the restoring footpoint symmetry process.

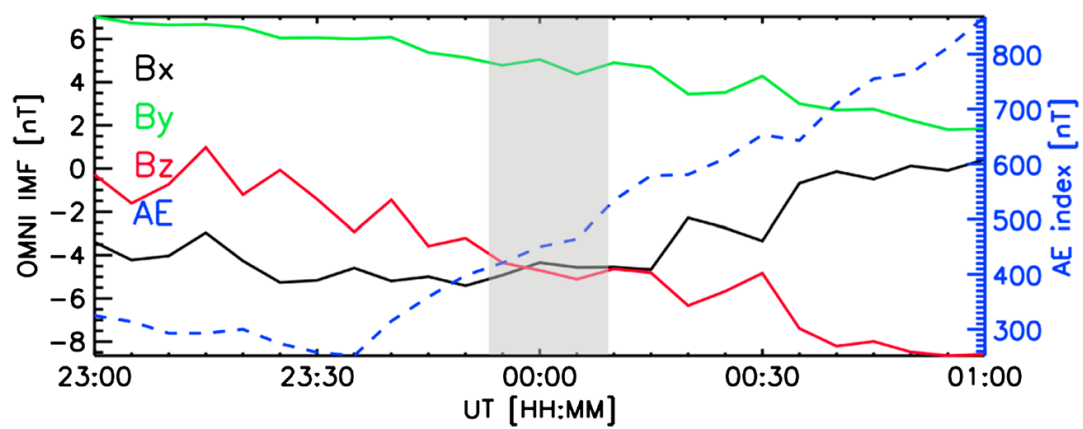

**Figure 3.** IMF $B_x$, $B_y$, and $B_z$ (solid) and the *AE* geomagnetic activity index (dashed) during the event on 18–19 May 2001. The shaded grey region indicates interval of simultaneous observations presented in Figure 5.

## 2. Instrumentation
### 2.1. Imaging Data
The auroral images from the Northern Hemisphere are obtained by the far ultraviolet (FUV) Wideband Imaging Camera (WIC) [*Mende et al.*, 2000] on board the IMAGE (Imager for Magnetopause-to-Aurora Global Exploration) spacecraft [*Burch*, 2000]. The WIC camera is sensitive to the UV range 140–190 nm, including the Lyman-Birge-Hopfield band (molecular nitrogen) and a few atomic nitrogen lines. The auroral emissions at these wavelengths are mainly due to precipitating electrons [*Frey et al.*, 2003]. The images are presented in modified apex coordinates (magnetic latitude (MLAT)/MLT) with reference height 130 km [*Richmond*, 1995]. In order to get an image only containing the aurora, emissions from dayglow as well as a constant noise level were subtracted.

The auroral images from the Southern Hemisphere are taken from the Polar spacecraft [*Acuña et al.*, 1995] VIS (Visible Imaging System) Earth camera [*Frank et al.*, 1995; *Frank and Sigwarth*, 2003]. The response in this camera is mostly from the OI line at 130.4 nm. The VIS Earth images are also presented in the same apex/MLT coordinates. As the auroral zone in the Southern Hemisphere was dark during the event, no dayglow removal was needed.

### 2.2. Ionospheric Plasma Flow Measurements
Measurements of the ionospheric plasma flow velocity are provided by two HF coherent scatter radars part of the Super Dual Auroral Radar Network (SuperDARN) [*Greenwald et al.*, 1995; *Chisham et al.*, 2007]. SuperDARN radars operate by transmitting radio signals that refract in the ionosphere and backscatter from decameter-scale, magnetic field-aligned irregularities in the electron density. Backscattered signals from the *F* region ionosphere experience a Doppler shift that is proportional to the line-of-sight (LOS) component of the plasma drift velocity. The two radars used in this study are the Stokkseyri radar, located in Iceland at geographic coordinates 63.86°N, 22.02°W, and with a bore site direction of 59.0°W, and the Syowa-East radar, located in Antarctica at 69.0°S, 39.58°E, and bore site 106.5°E. During the interval of interest, both radars were operating in a common mode in which they scan through 16 beams of separation 3.24°, with a total scan time of 2 min. Each beam is divided into 75 range gates of length 45 km, and so in each full scan the radars cover 52° in azimuth and over 3000 km in range.

## 3. Observations
In this section we present simultaneous auroral images and ionospheric plasma flow measurements around the nightside convection throat region in both hemispheres of an event starting on 18 May 2001.

IMF and geomagnetic activity around the event are presented in Figure 3. The interval of observations shown in this section is highlighted with grey shade. The IMF data are from the 5 min OMNI data product from NASA's Space Physics Data Facility [*King and Papitashvili*, 2005] and are represented in the geocentric solar magnetic (GSM) coordinate system. These data represent the IMF conditions at the Earth's bow shock nose. As the IMF shows little fluctuations during our event, any minor errors introduced by the timeshift is not important for our interpretation. Embedded in the OMNI data product is the $AE$ index (provided by the World Data Center for Geomagnetism, Kyoto), also with 5 min time resolution. Figure 3 shows that the IMF $B_z$ component declines slowly from ∼0 to −5 nT during the hour preceding the simultaneous ionospheric convection and imaging measurements 23:53–00:09 UT. During this time interval the IMF $B_x$ and $B_y$ components are stable with values around −5 and +5 nT, respectively. The $AE$ index is also shown in the same plot, indicating moderately disturbed geomagnetic conditions (∼450 nT). It should be noted that the longitudinal displacement of footpoint locations, $\Delta$MLT, is expected to depend on geomagnetic activity. In the statistical study of substorm onset locations in the two hemispheres by *Østgaard et al.* [2011b], the average hemispheric difference in MLT (∼ $\Delta$MLT) as a function of the IMF clock angle was found to be larger when only events with |IMF| > 5 nT were considered. This is expected as flux is added asymmetrically more efficiently for large values of IMF for the same clock angle. Although a more quiet event would be desirable to remove other disturbances, this event is unique in terms of data coverage to investigate the process of restoring symmetry and therefore represents a very interesting case. The 23:53–00:09 UT interval was chosen because of good radar coverage of the nightside convection throat region in the dusk cell.

### 3.1. Simultaneous Imaging of the Two Auroral Ovals

On 18 May 2001 the IMAGE FUV-WIC camera observed the Aurora Borealis from apogee, while the Polar VIS Earth camera simultaneously observed the Aurora Australis from perigee. This favorable configuration allowed simultaneous imaging of the auroral oval in both hemispheres.

A pair of simultaneous images plotted in the MLAT/MLT grid are seen in Figures 4a (Northern Hemisphere at 23:53:07 UT) and 4b (Southern Hemisphere at 23:52:55 UT). Note that the intensity scaling is chosen to best reflect the auroral variability within each hemisphere separately. The intensities cannot be directly compared between hemispheres due to the different instrumentation. Our interpretation is therefore limited to how relative differences within the images compare between the two hemispheres. The solid black curve in the Northern Hemisphere that is crossing midnight at around 70° MLAT is the contour of 90° solar zenith angle. It indicates that most of the aurora in the Northern Hemisphere is directly exposed to sunlight. However, as mentioned in section 2, the dayglow-related emissions have been removed. A distinct large-scale feature in the dusk side oval can be identified in both hemispheres, highlighted by the dashed line in Figures 4a and 4b. One can see that this feature is heavily displaced in MLT between the two hemispheres. The Southern Hemisphere part extends toward midnight, while it only extends toward 21 MLT in the Northern Hemisphere.

A correlation analysis is performed to more objectively identify the shift between the two hemispheres for the simultaneous image pairs. The images are first mapped to a rectangular grid. The VIS image (south) is held fixed, considering only the region from 18–03 MLT. The same region is extracted from the mapped WIC image including a variable shift in longitude (MLT). At 0.1 h steps, this MLT shift is varied from 0 to 4 h. Note that we use $\Delta$MLT = MLT$_{south}$ − MLT$_{north}$, resulting in a positive value of $\Delta$MLT for this case. Then $\Delta$MLT is identified as the relative shift at which the correlation between the extracted regions peaks. We use the linear correlation between the two vectors representing the fixed VIS image and the varying WIC image. This procedure is shown in Figures 4c and 4d. Here the same images as in the polar plots in Figures 4a and 4b are shown before and after the MLT shift is applied. This analysis will produce a value of $\Delta$MLT for each simultaneous image pair. In Figure 4e we show a time series plot of $\Delta$MLT. We have distinguished between good (black diamonds) and uncertain (grey diamonds) fits. This evaluation is based on visual inspection of the output for each image pair. Generally, the method works better when the aurora has similar structures in the two hemispheres. During the time interval when simultaneous ionospheric convection measurements were available in the conjugate regions, we derive a $\Delta$MLT as large as ∼3 h. To our knowledge, this is by far the largest MLT shift of conjugate footpoints that has been reported based on conjugate auroral observations.

We also note that $\Delta$MLT (Figure 4e) follows a similar trend as the IMF $B_y$ value (Figure 3), namely, that $\Delta$MLT is decaying as IMF $B_y$ decays. This is expected when the loading become more symmetric and consistent with statistical studies of auroral displacement and IMF $B_y$ [*Liou et al.*, 2001; *Wang et al.*, 2007; *Liou and Newell*, 2010; *Østgaard et al.*, 2011b]. We also note that there is a broad peak of $\Delta$MLT from about 2350 to 0030 UT of about 1 h more than the general decrease of IMF $B_y$ implies. Although we believe that the large displacement seen in this event is due to the combined effect of positive IMF $B_y$ and positive dipole tilt [*Liou and Newell*, 2010], we have no explanation for this broad peak. We emphasize that the restoring symmetry process occurs as closed field lines convect from the tail toward the dayside and footpoint symmetry is restored ($\Delta$MLT$\rightarrow$ 0 for that particular field line). This is different from the situation when $\Delta$MLT$\rightarrow$ 0 on the nightside, which is related to the driving mechanism disappearing (IMF $B_y \rightarrow 0$). During a constant IMF $B_y$ interval, as shown in this paper, $\Delta$MLT should stay the same in the nightside region.

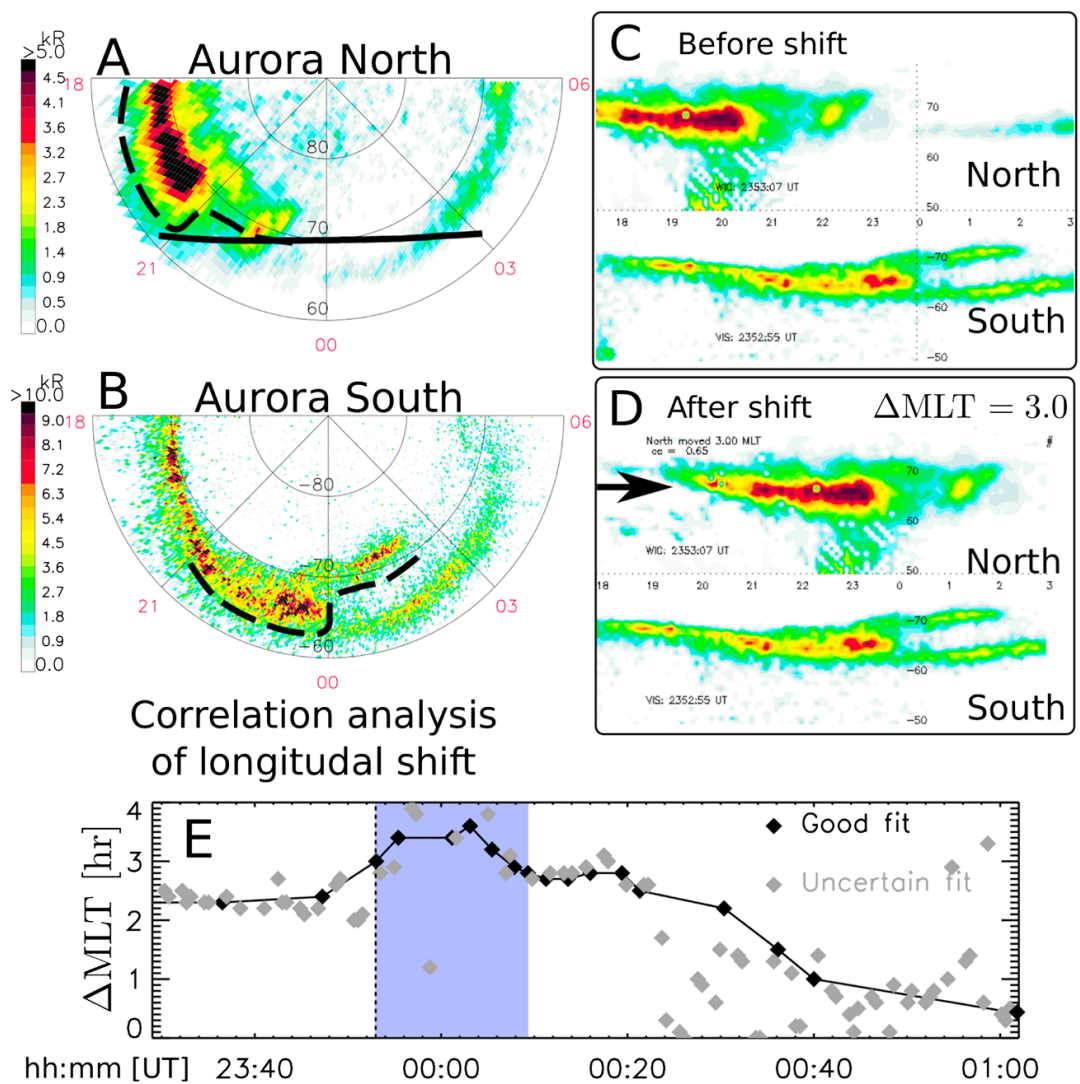

**Figure 4.** (a and b) Simultaneous images from WIC in the Northern Hemisphere and VIS in the Southern Hemisphere on a polar MLAT/MLT grid. (c) The same images projected on a rectangular MLAT/MLT grid before a longitudinal shift is applied. (d) Same as Figure 4c but now the Northern Hemisphere image has been shifted 3 h in the direction indicated by the arrow. (e) Time series of the longitudinal shift of the aurora between the hemispheres derived from a correlation analysis. Black diamonds indicate an accurate determination. The vertical dashed line indicates the time of the image pair presented, and the blue shaded region indicate the time interval when conjugate convection from SuperDARN is analyzed in Figures 5 and 7.

### 3.2. Simultaneous Ionospheric Convection and Hall Currents

Two of the SuperDARN radars were favorably located during this event to monitor the plasma flow near the nightside convection throat. The LOS plasma flow velocity is plotted on top of the auroral images for four instances in Figures 5a–5d, with toward velocities as red and away velocities as blue. Of special interest is the nightside convection throat, which in the framework of the restoring symmetry concept needs to be a conjugate feature. Upon successful identification in both hemispheres, this will provide an independent measure of $\Delta$MLT. We identify the nightside convection throat as the transition from away to toward velocities in the radar data when the LOS direction is in the east/west direction and echoes originate from closed field lines.

In the Southern Hemisphere, the Syowa East radar observed the nightside convection throat. The radar was located around magnetic midnight and pointing in a magnetic southeast direction. In Figure 5 the reversal is seen clearly at 23:53 UT around 02 MLT (Figure 5a, bottom). Here the transition between toward and away plasma flow occurs along a distance of only $\sim$200 km in the LOS direction. There is also a small uncertainty regarding the actual location of SuperDARN echoes due to mapping errors [e.g., *Yeoman et al.*, 2001, 2008; *Chisham et al.*, 2008]. These studies found the mapping error at distances similar to those considered here ($\sim$1000 km and less) to be typically less than 30 km, which is smaller than one radar range gate (45 km). Hence, we claim that the Southern Hemisphere convection throat region at 23:53 UT is accurately determined to be $02 \pm 0.1$ MLT at 70° MLAT. We have indicated the shape of the extended dusk cell as the thick dashed curve in Figure 5a (bottom). In the later image pairs shown in Figures 5b–5d, the Southern Hemisphere convection throat is not captured as accurately due to the decreasing number of away echoes. But since the plasma flows remain westward, it indicates that 02 MLT is the minimum dawnward extent of the dusk cell.

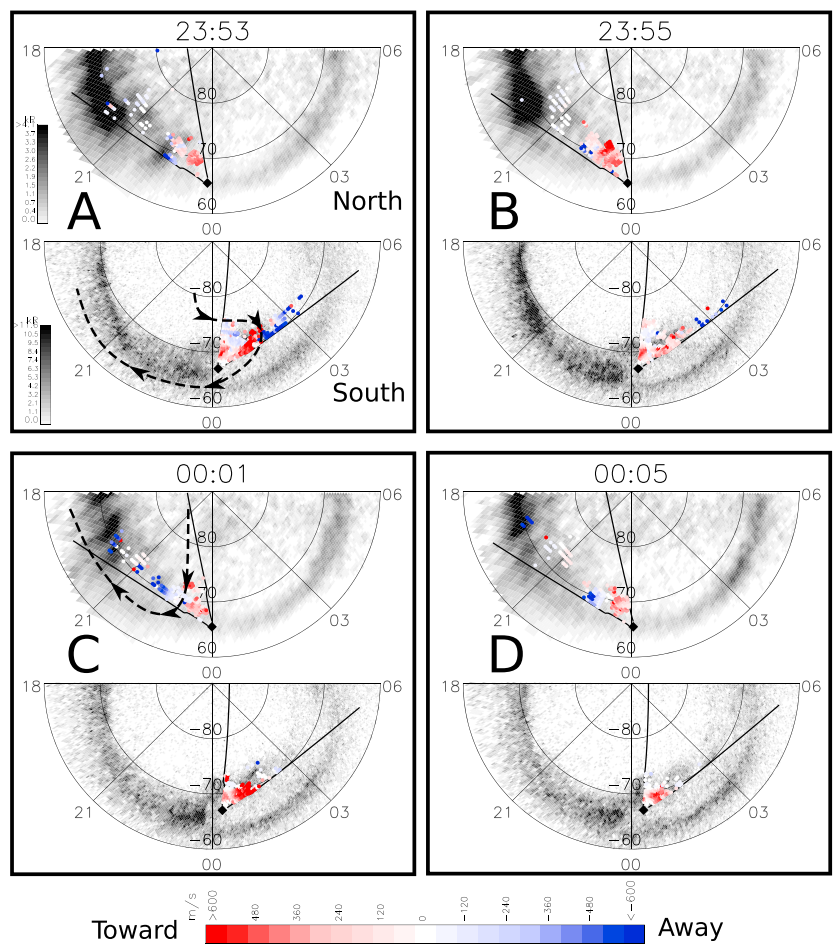

**Figure 5.** (A-D) SuperDARN line-of-sight velocities around the nightside reversal region plotted on top of the simultaneous conjugate auroral images at four instances during the event. Plasma flows toward the radar are shown in red; plasma flows away are shown as blue. Based on these radar data, the likely extent of the dusk convection cells is drawn with dashed lines in both hemispheres. One can see that the nightside convection throat is shifted $3.2 \pm 0.6$ h in MLT between the hemispheres, consistent with the auroral observations.

The Stokkseyri radar in the Northern Hemisphere was located at magnetic midnight and pointing in a magnetic northwest direction. Its field of view and location are indicated by the black lines originating from the black diamond in the Northern Hemisphere plots in Figures 5a–5d (top). In the Northern Hemisphere, the nightside convection throat region is best captured at 00:01 UT, see Figure 5c (top). Here the equatorward three beams indicate a transition from plasma flows away from the radar (blue) at 22.3 MLT, through a region of small LOS magnitudes, indicating the plasma flow is primarily perpendicular to the LOS direction, to plasma flows toward the radar at 23.3 MLT. Based on this, we have sketched a possible extent of the dusk convection cell at this instance, seen as the thick dashed line in Figure 5c. From these data, the nightside convection throat in the Northern Hemisphere is likely within $22.8 \pm 0.5$ MLT.

We present further evidence for the nightside convection throat location in the Northern Hemisphere in Figure 6. Here magnetic perturbations from ground magnetometers from the SuperMAG network [*Gjerloev*, 2012] are shown on a MLT/MLAT map at 00:01 UT. The perturbation vectors are rotated 90° clockwise to represent the ionospheric equivalent current. Due to the high conductivity from sunlight in this hemisphere, the magnetic perturbations are likely dominated by the ionospheric Hall current overhead [*Laundal et al.*, 2015] and are therefore antiparallel to the ionospheric convection. In the auroral region, (below 70° MLAT) a transition from eastward Hall currents (seen at the Greenland NAQ station at 22.4 MLT) to westward Hall currents (seen as far west as the Iceland LRV station at 0.0 MLT) is observed, corresponding to the two cell convection pattern. This transition is centered in the same MLT region as the SuperDARN data shown in Figure 5c, strongly suggesting that the signature we see in the radar data likely is the nightside convection throat region. The SuperMAG data show only minor variations in the region of interest during the time window presented in Figure 5, indicating that the convection throat remained around 23 MLT in the Northern Hemisphere.

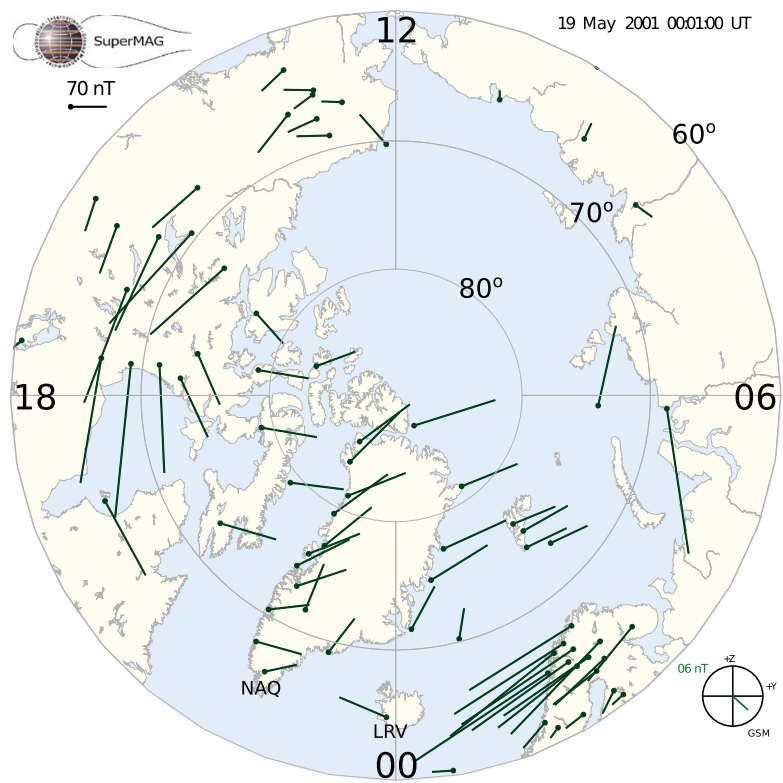

**Figure 6.** Ground magnetic perturbations from the SuperMAG network [*Gjerloev*, 2012] rotated 90° clockwise on a MLT/MLAT grid, representing the ionospheric Hall currents. The two-cell pattern in the Hall currents changes direction between Greenland and Iceland in agreement with our interpretation of the SuperDARN data.

When comparing the nightside convection throat location in the two hemispheres, we arrive at a very similar longitudinal displacement between hemispheres as seen in the aurora in Figure 4. Due to the uncertainty in defining the transition region from the radar data, we conclude that the longitudinal displacement of the nightside convection throat is within $3.2 \pm 0.6$ h MLT during the interval presented.

In an effort to compare the velocities more directly in conjugate regions (westward of the nightside convection throat in the dusk cell), we present averaged convection velocities as a function of time in Figure 7. We consider only the return convection in the dusk cell in the vicinity of the nightside convection throat. For the Northern Hemisphere we consider echoes equatorward of 73° MLAT and closer than 21 MLT that indicate plasma flow away from the radar (blue). For these echoes, the LOS velocities are less than 10° off the apex westward direction. The average of these echoes is plotted as the blue line in Figure 7 with its corresponding standard error. The same is done in the Southern Hemisphere. Here we only consider the echoes that indicate convection toward the radar (red) in the five most equatorward beams at ranges equatorward of −73° MLAT. In this way the LOS velocity are mostly within 20° of the apex westward direction, and the average velocity represents the westward convection on closed field lines. This is plotted in red with its corresponding standard error in Figure 7. A persistent stronger westward convection is seen in the Southern Hemisphere compared to its conjugate region in the Northern Hemisphere. This is in qualitative agreement with the expected behavior (from restoring symmetry process) in the dusk cell when ΔMLT is positive, as described in section 1.

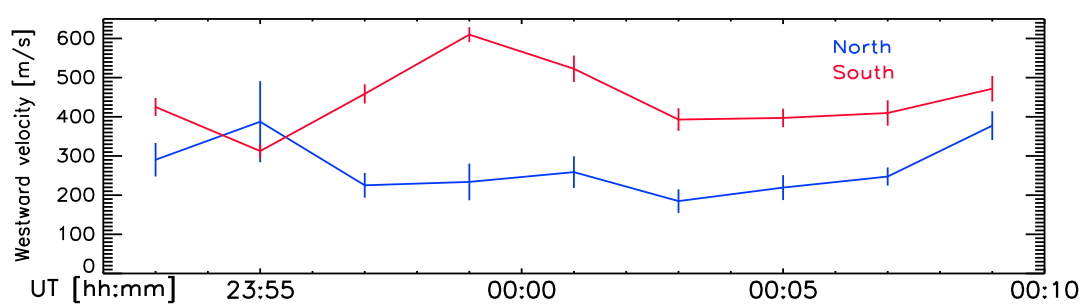

**Figure 7.** Averaged westward velocities in the dusk cell on closed field lines in the two hemispheres. A persistent stronger westward convection is seen in the Southern Hemisphere compared to its conjugate region, consistent with the restoring symmetry process.

## 4. Discussion

### 4.1. Displaced Footpoints and Asymmetric Convection

From the direct comparison presented in Figure 7 we conclude that the westward ionospheric convection on closed field lines close to the nightside convection throat region in the dusk cell is stronger in the Southern Hemisphere compared to its conjugate region. This difference in ionospheric plasma flow velocity is expected for field lines with asymmetric footpoints on the nightside with positive ΔMLT, and it is consistent with earlier observations [*Grocott et al.*, 2005, 2007; *Pitkänen et al.*, 2015]. It is also consistent with the concept of restoring symmetry described in section 1 and by *Tenfjord et al.* [2015], where the release of magnetic stress into one hemisphere causes plasma to convect faster on the equatorward part of the banana cell. It should be noted that the westward plasma flow values observed here are less intense than the fast plasma flows (∼1000 m/s) reported by, e.g., *Grocott et al.* [2004]. However, individual echoes indicate velocities up to ∼800 m/s in Figure 5. The discrepancies between northern and southern convection velocities are nevertheless attributed the restoring symmetry process as the sign of IMF $B_y$, the large ΔMLT, and the observed plasma flow direction are all consistent.

The earlier observations [*Grocott et al.*, 2005; *Pitkänen et al.*, 2015] focused on the fast plasma flows related to the restoring symmetry process, being on opposite ionospheric convection cells in the two hemispheres. Consequently, they did not focus on the plasma flow at its conjugate location as we do in this paper. Hence, identification of a longitudinal displacement of conjugate points is uncertain and difficult in their cases, making it less clear that their observed fast convection was really asymmetric at its conjugate location. However, in our case we also have simultaneous conjugate imaging in both hemispheres. Our firmly established footpoint asymmetry together with the observed hemispheric difference in convection in conjugate regions demonstrate that conjugate regions are affected differently, as expected from the influence of the restoring symmetry process. We also note that this is to our knowledge the largest value of ΔMLT reported from conjugate auroral imaging. We suggest that the large value is related to the combined effect of positive IMF $B_y$ and a positive dipole tilt. As shown in Figure 3 of *Liou and Newell* [2010], $B_y$ related to warping of the plasma sheet and the induced $B_y$ from the IMF will add in the premidnight sector for this situation, possibly contributing to the large value of ΔMLT observed here.

### 4.2. Effect on the Aurora

Having simultaneous conjugate imaging could in principle allow us to examine the impact of the restoring symmetry process on auroral brightness. Since the restoring symmetry process is proposed to be associated with stronger BCs in the hemisphere of faster convection (corresponding to the banana convection cell, see Figure 2), this could also have an effect on auroral brightness, as the aurora seen by the UV imagers is mostly due to accelerated precipitating electrons associated with upward current regions [*Waters et al.*, 2001; *Paschmann et al.*, 2002; *Mende et al.*, 2003a, 2003b; *Dubyagin et al.*, 2003]. Comparing auroral intensities between hemispheres is by no means straightforward in this case. The two cameras are sensitive to emission lines from different atmospheric species (oxygen and nitrogen). For different seasons, the atmospheric composition varies, affecting the relative intensity between the emissions observed by VIS and WIC [*Laundal and Østgaard*, 2009]. Also, the viewing geometry and the area observed in each pixel differ between hemispheres making the direct comparison challenging. However, relative differences within the images from the two hemispheres can be identified with better confidence. An example of this is the poleward structure seen in the Southern Hemisphere from 00 to 02 MLT in Figure 4b. One can see that it has an intensity similar to the main oval in the dusk sector in the same hemisphere. The same poleward structure is also seen in the Northern Hemisphere from 21 to 23 MLT in Figure 4a. Here the intensity is less than the main dusk oval in the same hemisphere. Hence, this poleward structure appears more distinct in the Southern Hemisphere than its conjugate counterpart. This can be expected from stronger BCs associated with the restoring symmetry process we observe in the dusk cell in this region. However, we cannot rule out other mechanisms responsible for the observed asymmetry, such as parallel potential drops only in the dark hemisphere [*Newell et al.*, 1996]. To further investigate the suggested influence on BC from the restoring symmetry process, we will in the following look directly at BC signatures rather than associated auroral emissions.

### 4.3. Birkeland Currents Associated With the Restoring Symmetry Process

As we do not have any measurements of the BCs in the region where the plasma flows associated with the restoring symmetry process are seen in the presented event, we have made an effort to investigate the average large-scale BC distribution during the same conditions as in our event. This is done using the Active Magnetosphere and Planetary Electrodynamics Response Experiment (AMPERE) [*Anderson et al.*, 2000] with data from January 2010 to May 2013. Based on global sampling of the magnetic field perturbations at 700 km by the Iridium satellites, global instantaneous maps of BCs from both hemispheres are derived every 2 min based on a 10 min sliding average window [*Waters et al.*, 2001].

We present average maps of BCs from both hemispheres during conditions similar to the event presented in section 3. We select only observations having the IMF $B_y$ and $B_z$ values in the range that was observed during the hour prior to our event, being [5, 7] nT for IMF $B_y$ and [−5, 0] nT for IMF $B_z$. The IMF values were calculated using a 50 min average (from 40 min prior to 10 min later) of the 5 min OMNI IMF to allow the closed magnetosphere to reconfigure. This long averaging interval combined with the large average value (5–7 nT) will ensure that the IMF $B_y$ forcing has been substantial prior to the selected measurements. To ensure stability of IMF $B_y$, we also require that all individual 5 min values are within the range 4–8 nT for each event. Furthermore, we only include observations from the same season as the event. This is done by selecting only AMPERE observations when the dipole tilt is greater than 10° (dipole tilt is defined as positive for northern summer/southern winter). We also use the stability criteria defined by *Anderson et al.* [2008] to avoid periods when the Birkeland current pattern is rapidly changing. The stability criteria are a measure of relative overlap between consecutive patterns, quantified by a coefficient between 0 and 1. We use the same threshold value as *Anderson et al.* [2008], 0.45, but since the AMPERE data are now available at higher time resolution, our coefficients are based on current maps separated by 20 min, rather than the 1 h used by *Anderson et al.* [2008]. Selecting only stable current patterns for the averaging ensures that the current has had time to reconfigure in response to the IMF orientation. Furthermore, we only present observations from the nightside (18–06 MLT) as this is the region of interest for this paper. The number of individual current maps used in the statistics is indicated in each subplot, as well as the maximum and minimum average current density value in the premidnight and postmidnight regions separately (printed in red and blue).

The expected BCs related to the restoring symmetry process are sketched in Figure 2. It suggests that for IMF $B_y$ positive, the Southern Hemisphere dusk side should experience a pair of upward/downward BCs at a poleward/equatorward location, respectively. This is the same polarity as the large-scale region 1/region 2 (R1/R2) pattern. Hence, one would expect to see an enhancement of both. Applying the same reasoning to a closed field line on the dawn cell, the Northern Hemisphere dawn side R1/R2 should also be enhanced during IMF $B_y$ positive. In Figure 8a we show the average nightside BC maps from both hemispheres representing the large-scale BC pattern during the event presented in section 3 when plasma flows associated with the restoring symmetry process are seen. An apparent dawn-dusk asymmetry in current density can be seen in Figure 8a, consistent with the proposed influence from the restoring symmetry process, namely, stronger Southern Hemisphere dusk cell BCs and stronger Northern Hemisphere dawn cell BCs. The differences are strongest in the R1 currents especially in Northern Hemisphere, but also seen in the R2 currents. Also, a ∼3 h ΔMLT is seen from the extent of the R1/R2 pairs between hemispheres (indicated with the black dashed line in Figure 8a), in good agreement with both auroral and nightside convection throat observations during the event. We suggest that this large average rotation of the nightside R1/R2 current pattern for these specific conditions is related to the combined effect of positive dipole tilt and positive IMF $B_y$ as shown in Figure 3 of *Liou and Newell* [2010]. For opposite conditions (negative dipole tilt and negative IMF $B_y$) the average AMPERE BC pattern resemble mirror images of Figure 8a indicating a similar (but opposite) value for ΔMLT. This is shown in Figure 8c where the same IMF criteria (but opposite sign of IMF $B_y$) are used. We also show the case when both IMF $B_y$ and dipole tilt are small (less than 2 nT and 5°, respectively) in Figure 8b, indicating no major dawn-dusk asymmetries in the BC strength and also no significant ΔMLT in the nightside, as expected when there is no loading ($B_y$) or warping of the neutral sheet (tilt). The effect on BCs of increasing versus decreasing IMF $B_y$ was also investigated by separating the events (selected without the IMF stability criterion) into the two categories (not shown). No significant difference in the average BCs could be seen between the two categories.

In Figures 8a and 8c it is evident that signatures of the restoring symmetry process is not only seen in the immediate vicinity of the nightside convection throat but extends toward the dayside (here only shown toward 18 and 06 MLT). If these signatures are related to the restoring symmetry process, it indicates that this process is present along a large portion of the return convection region toward the dayside. From the MHD modeling study by *Tenfjord et al.* [2015], it was shown that $B_y$ is induced also on the dayside closed field lines in a similar manner (same direction as IMF $B_y$). Their analysis indicated that this also affects the footpoint locations, leading to a negative ΔMLT on the dayside in response to a positive IMF $B_y$. This suggests that the initial displaced field lines on the nightside (positive ΔMLT) will be displaced in the opposite direction (negative ΔMLT) before arriving at the dayside, pointing to a location of ∼0 ΔMLT somewhere in the dusk (dawn) sector, likely centered around 18 (06) MLT. The location of 0 ΔMLT on the dusk side can be estimated for the event studied in this paper by using the observed convection velocities in the nightside convection throat region presented in Figure 7, assuming that the velocities stay constant during the restoring symmetry process. Assuming that the Southern Hemisphere footpoint moves 300 m/s faster at 70° MLAT, it will take 95 min to catch up with the 3 h ΔMLT. During that time the Northern Hemisphere end will have moved 3 h in MLT at 300 m/s. This suggests a total extent of ∼ 6 h in MLT of the restoring symmetry process, in reasonable agreement with the predicted location at ∼18 MLT. Such a large extent of the restoring symmetry process is in agreement with the large region of dawn/dusk asymmetries in BCs seen in Figure 8, hence supporting the understanding of the dynamic effects of restoring magnetic footpoint symmetry.

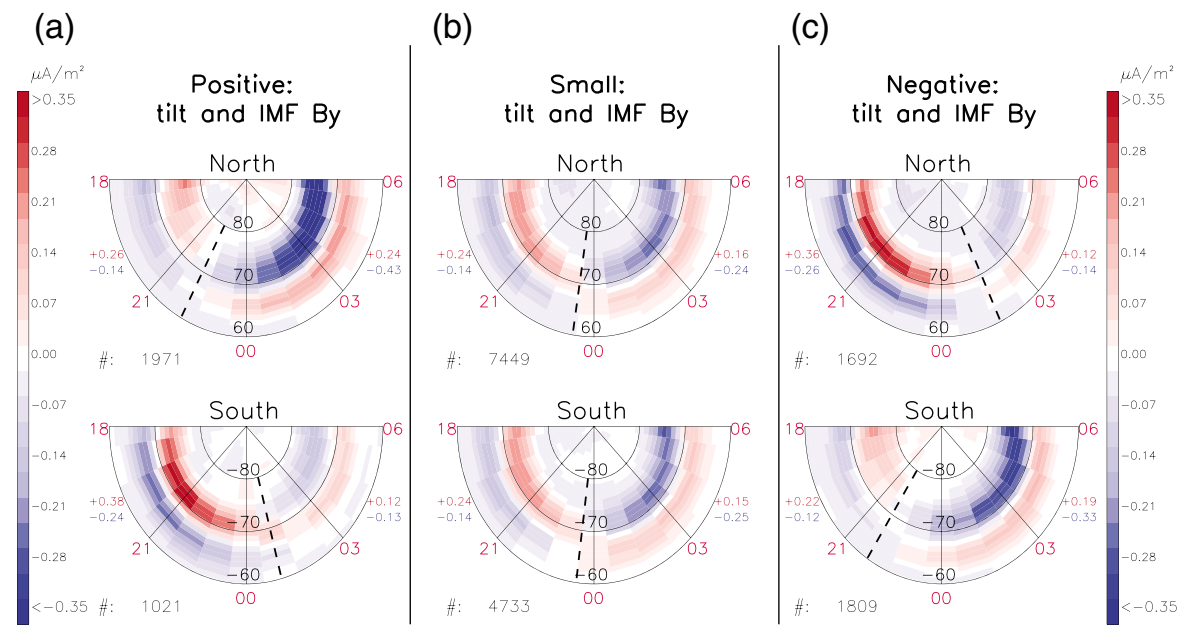

**Figure 8.** (a) Average maps of BC from AMPERE corresponding to the event presented in section 3 (positive IMF $B_y$ and negative IMF $B_z$). Northern summer and southern winter is presented by selecting observations only when dipole tilt >10°. Dawn-dusk asymmetries in current density is seen, consistent with the expected influence from the restoring symmetry process. The number of individual current maps is indicated in each subplot, as well as the maximum and minimum average current density value both premidnight and postmidnight (printed in red and blue). The dashed black line indicate the approximate location where the R1/R2 currents change polarity, suggesting a ΔMLT ∼ 3 h also from the AMPERE data. (b) Similar average BC maps during small (< 2 nT) IMF $B_y$ and small (<5°) dipole tilt showing no major differences between the two hemispheres. (c) Same as Figure 8a but for negative IMF $B_y$ and negative dipole tilt. The average dawn-dusk asymmetry and rotation of current pattern appear to be a result of the combined effect of dipole tilt and IMF $B_y$, consistent with observations by *Liou and Newell* [2010] and *Petrukovich* [2011]. This asymmetry is in qualitative agreement with influence from the restoring symmetry process.

Since IMF $B_y$ is known to lead to asymmetric footpoints on closed field lines in the nightside, the restoring symmetry process might be of importance in a more general sense. It is therefore of interest to look at previous relevant work during conditions favorable for the restoring symmetry process to occur.

*Green et al.* [2009] used the Iridium satellites to investigate the global BC response to IMF and season. Although their selection criteria were less strict regarding the IMF $B_y$ magnitude than we use in Figure 8, their local winter maps show the same trend as discussed for Figure 8 (but less distinct) regarding a dawn-dusk intensity asymmetry of the BC pattern related to IMF $B_y$. However, they did not comment on this feature.

The MFACE (Model of Field-Aligned Currents using Empirical orthogonal functions) empirical BC model based on CHAMP magnetic field data [*He et al.*, 2012] also captures a dawn-dusk asymmetry in the R1/R2 current pattern depending on IMF $B_y$. Although the signatures in their static model are weak [*He et al.*, 2012, Figure 3], they are consistent between hemispheres for both signs of IMF $B_y$.

Using ground magnetometers in the Northern Hemisphere, *Friis-Christensen and Wilhjelm* [1975] observed a stronger westward electrojet during winter conditions for positive IMF $B_y$ compared to negative IMF $B_y$. They did comment on this feature, only to conclude that it was most likely not an observational bias and that it most likely was related to IMF $B_y$. We suggest that the increased westward electrojet they observed due to IMF $B_y$ could be related to increased BCs in the dawn cell, consistent with the restoring symmetry process. Influence

**Acknowledgments**
We thank S.B. Mende and the IMAGE FUV team at the Space Sciences Laboratory at UC Berkeley for the WIC data. The WIC images were processed using the FUVIEW3 software (http://sprg.ssl.berkeley.edu/image/). We thank Rae Dvorsky and the Polar VIS team at the University of Iowa for the VIS Earth data. The VIS Earth images were downloaded from NASA's Space Physics Data Facility (ftp://cdaweb.gsfc.nasa.gov/pub/data/polar/vis/vis_earth-camera-full/) and processed using the XVIS 2.40 software (http://vis.physics.uiowa.edu/vis/software/). We acknowledge the use of NASA/GSFC's Space Physics Data Facility (http://omniweb.gsfc.nasa.gov) for OMNI data. We also thank Brian Anderson and the AMPERE team for sharing their data. Operation of the SuperDARN radars is supported by the national funding agencies of the U.S., Canada, the U.K., France, Japan, Italy, South Africa, and Australia. For the ground magnetometer data we gratefully acknowledge: Intermagnet; USGS, Jeffrey J. Love; CARISMA, PI Ian Mann; CANMOS; The S-RAMP Database, PI K. Yumoto and Dr. K. Shiokawa; The SPIDR database; AARI, PI Oleg Troshichev; The MACCS program, PI M. Engebretson, Geomagnetism Unit of the Geological Survey of Canada; GIMA; MEASURE, UCLA IGPP and Florida Institute of Technology; SAMBA, PI Eftyhia Zesta; 210 Chain, PI K. Yumoto; SAMNET, PI Farideh Honary; The institutes who maintain the IMAGE magnetometer array, PI Eija Tanskanen; PENGUIN; AUTUMN, PI Martin Connors; DTU Space, PI Dr. Juergen Matzka; South Pole and McMurdo Magnetometer, PI's Louis J. Lanzarotti and Alan T. Weatherwax; ICESTAR; RAPIDMAG; PENGUIn; British Artarctic Survey; McMac, PI Dr. Peter Chi; BGS, PI Dr. Susan Macmillan; Pushkov Institute of Terrestrial Magnetism, Ionosphere and Radio Wave Propagation (IZMIRAN); GFZ, PI Dr. Juergen Matzka; MFGI, PI B. Heilig; IGFPAS, PI J. Reda; University of LAquila, PI M. Vellante; SuperMAG, PI Jesper W. Gjerloev. This study was supported by the Research Council of Norway under contract 223252 and the Peder Sather Center for Advanced Study. A.G. was supported by Science and Technology Facilities Council (STFC) grant ST/M001059/1. H.U.F. was supported by NSF through the grant GIMNASTAGS-1004736. S.E.M. was supported by the STFC, UK, grant ST/K001000/1.

on ground magnetometers from BCs has recently been suggested to be important when the conductivity is low [*Laundal et al.*, 2015].

The material presented in this paper indicates that the large-scale behavior seen in the nightside by asymmetric footpoints (Figure 4), asymmetric plasma flows (Figure 7), and asymmetric BCs (Figure 8a) can be explained in the framework of the restoring symmetry process as described in section 1 and by *Tenfjord et al.* [2015]. We emphasize that our description of the dynamic restoring symmetry process is broadly consistent with the electrostatic description of the large-scale convection pattern, where the ionospheric convection electric field can be expressed as the gradient of a potential, $\vec{E} = -\nabla\Phi$. This is expected since the average BC patterns presented in Figure 8 represent a steady state. In the electrostatic description, the ionospheric electric potential and BCs are related through $j_\parallel \sim \nabla^2\Phi$ when neglecting conductivity gradients. This is often further simplified to $j_\parallel \sim \nabla \times \vec{v}$ [*Sofko et al.*, 1995] where $\vec{v}$ is the convection field. It is therefore not surprising that the observed dawn-dusk asymmetries in the BC pattern are in good agreement with the differently shaped (different vorticity) convection cells in dawn and dusk known to be largely controlled by IMF $B_y$. What is new, however, is the dynamic description of how this situation is first established and then maintained. This is what we explain as the asymmetric loading process (described in the introduction and by *Tenfjord et al.* [2015]) and then the restoring symmetry process, affecting closed field lines in the nightside portion of the two convection cells. Hence, our main conclusion is to recognize that these presented observations are likely a result of the dynamic effects of restoring footpoint symmetry on closed field lines in the nightside.

## 5. Conclusion

We have presented an event where a large longitudinal shift of the aurora between the hemispheres is seen. This is interpreted as evidence of closed field lines in the nightside having very asymmetric footpoints associated with the persistent positive IMF $B_y$ before and during the event. This large hemispheric shift of field line footpoint, being 3 h in MLT, is to our knowledge the largest value of ΔMLT reported on from conjugate auroral imaging. Simultaneous ionospheric convection measurements in the nightside convection throat region indicate that the conjugate footpoints respond to the asymmetric configuration set up by the solar wind-magnetosphere interaction by imposing faster convection in one hemisphere, consistent with the restoring symmetry process described here.

Average BC patterns during similar conditions as the event are shown. It indicates that the event under consideration experienced dawn/dusk asymmetries in the large-scale BCs consistent with the expected influence from the restoring symmetry process.

The presented material is interpreted as evidence that the asymmetric footpoint, the asymmetric plasma flows, and the asymmetric BCs all can be explained in the framework of the restoring symmetry process. Hence, our main conclusion is to recognize that these presented observations are likely a result of the dynamic effects of restoring footpoint symmetry on closed field lines in the nightside.